# Terahertz-infrared dielectric properties of lead-aluminum double-cation substituted single-crystalline barium hexaferrite

Asmaa Ahmed [a,b,*], Anatoly S. Prokhorov [a,c], Vladimir Anzin [a,c], Denis Vinnik [d], Sergei Ivanov [e], Adam Stash [f], Y.S. Chen [g], Alexander Bush [h], Boris Gorshunov [a], Liudmila Alyabyeva [a]

[a] Laboratory of Terahertz Spectroscopy, Center for Photonics and 2D Materials, Moscow Institute of Physics and Technology, 9 Institutskiy per., Dolgoprudny, Russia

[b] Department of Physics, Faculty of Science, Sohag University, Sohag University St, Sohag, Egypt

[c] Prokhorov General Physics Institute of the Russian Academy of Sciences, 38 Vavilov St, Moscow, Russia

[d] South Ural State University, 76 Lenin Prospect, Chelyabinsk, Russia

[e] Moscow State University, Chemistry Department, Moscow, Russia

[f] Nesmeyanov Institute of Organoelement Compounds Russian Academy of Sciences, 28 Vavilov St, Moscow, Russia

[g] NSF's ChemMatCARS Beamline@APS, The University of Chicago, Argonne, IL, 60439, USA

[h] Research Institute of Solid-State Electronics Materials, MIREA − Russian Technological University (RTU MIREA), 78 Vernadsky prospect, Moscow, Russia

* Corresponding author,
E-mail address: a.gamal@phystech.edu (A. Ahmed)

**Abstract**

Hexaferrite materials are highly demanded to develop and manufacture electronic devices operating at radio- and microwave frequencies. In the light of the prospects for their use in the forthcoming terahertz electronics, here, we present our results on the terahertz and infrared dielectric response of a typical representative of hexaferrites family, lead-substituted M-type barium hexaferrite doped with aluminum, $Ba_{0.2}Pb_{0.8}Al_xFe_{12-x}O_{19}$, $x$(Al)=0.0, 3.0, and 3.3. We studied uniquely large and high-quality single crystals of the compounds prepared by spontaneous crystallization growth technique. Our aim was to explore the effect of aluminum substitution on the dielectric response of the compounds. Systematic and detailed investigations of the dependences of terahertz-infrared (frequencies 8 - 8000 $cm^{-1}$) spectra of complex dielectric permittivity on the temperature, 4 - 300 K, and on the chemical composition, $x$(Al)=0.0, 1.2, 3.0, 3.3, were performed for two principal polarizations of the electric field **E**-vector of the probing radiation relative to the crystallographic $c$-axis, namely **E**$\|c$ and **E**$\perp c$. Furthermore, infrared phonon resonances are recorded and discussed. In contrast to undoped $BaFe_{12}O_{19}$, no softening of the lowest frequency $A_{2u}$ phonon is observed, indicating suppression of a displacive phase transition in substituted compounds. A number of resonance absorption bands are discovered at terahertz frequencies and assigned to transitions between energy levels of the fine-structured ground state of $Fe^{2+}$ ($^5E$) ions. The temperature and aluminum-doping dependences of the resonances are analyzed with an account taken of disorder introduced by aluminum. Basing on dielectric data and detailed X-ray experiments, we find that for all concentrations of $Al^{3+}$ ions, $x$(Al)=0.0, 1.2, 3.0, and 3.3, they mainly occupy the 2$a$ and 12$k$ octahedral site positions and that the degree of substitution of iron in tetrahedral positions is not substantial. Along with fundamental findings, the obtained data on broad-band dielectric properties of $Ba_{0.2}Pb_{0.8}Al_xFe_{12-x}O_{19}$ crystals provides the information that can be used for development and manufacture of electronic devices with operating frequencies lying in the terahertz spectral band.

1. **Introduction**

M-type hexaferrites with general formula $A^{2+}Fe_{12}O_{19}$, hereafter called BaM, PbM, SrM ($A^{2+}$ is a divalent cation $Ba^{2+}$, $Pb^{2+}$, $Sr^{2+}$, etc.), have received a great deal of attention since their discovery over 70 years ago. BaM has distinct physical properties that make it promising for applications in various areas, including permanent magnets, microwave devices, absorbers, high-quality magnetic memory storage elements, etc. [1–9]. Functional properties of hexaferrites are strongly susceptible to doping that can be realized by substituting $Ba^{2+}$ or $Fe^{3+}$ or both with other elements under the condition to keep the charge balance. Aluminum is one of the promising dopants in the hexaferrites family as it strongly affects various characteristics of the compounds – crystal field, anisotropy fields, ferromagnetic resonance frequency, Curie temperature [10–14]. For example, at relatively low concentrations of Al, the coercive force is small, while it grows up significantly at higher Al concentrations [15–17]. In addition, the increased concentration of Al causes suppression of the Curie temperature of the composites [2] [13]. It should be noted that the available experimental data on M-type hexaferrites are focusing mainly on the crystal structure, preparation techniques and magnetic properties of ceramic forms of the material (see [18–20] and references therein), and there are only a few recent investigations on the dielectric properties, especially at terahertz and infrared frequencies, of single-crystalline forms of M-type hexaferrites [21,22]. At the same time, these compounds, which already find wide applications in microwave electronics [21,29,30], are of increasing interest from the viewpoint of use in the next-generation electronic devices that will operate at as high as terahertz frequencies. The present study aims at filling up the gap by detailed investigating the influence of substitution of $Fe^{3+}$ with $Al^{3+}$ on the terahertz and infrared (THz, IR) dielectric properties of the material by measuring the broad-band spectra of complex dielectric permittivity $\varepsilon^* = \varepsilon' + i\varepsilon''$ of $Ba_{0.20}Pb_{0.80}Fe_{12-x}Al_xO_{19}$ in a wide range of frequencies $v$=0.2 THz – 240 THz ($\approx$ 8 - 8000 $cm^{-1}$) and at temperatures in the interval $T$ = 4 – 300 K. To analyze the disorder induced by Pb-Ba substitution, we provide structure refinement of the $PbFe_{12}O_{19}$ compound.

2. **Materials and Methods**

Single crystals of $Ba_{0.2}Pb_{0.8}Al_xFe_{12-x}O_{19}$ [$x$(Al)=0.0, 3.0, and 3.3] with the lateral dimensions up to 5 mm were grown using spontaneous crystallization growth technique. In the synthesis process, PbO, $B_2O_3$, $Fe_2O_3$, $Al_2O_3$, and $BaCO_3$ were used as initial ferrites components. During growth, iron and aluminum oxides were mixed in a 2: 1 molar ratio. The mixture of PbO and $B_2O_3$ was used as a flux. The charge includes up to 80 wt.% of the flux. The mixtures were ground in an agate mortar and filled into a 30 ml platinum crucible. The crucible was placed in a furnace. To homogenize the components, the melt was heated to 1533 K for three hours. Within 140 hours, the temperature was decreased to 1173 K with a speed of 5 K/h. The crystals with hexagonal shapes were separated from the flux by leaching in hot nitric acid.

The chemical compositions of the grown crystals were determined using an electron microscope Jeol JSM-7001F with the energy dispersive X-ray fluorescence spectrometer Oxford INCA X-max 80 (SEM/EDX). The phase composition was determined using powder X-ray diffractometer Rigaku Ultima IV(PXRD). For powder X-ray diffraction and elemental analysis, several well faceted single crystals were ground to obtain averaged results.

The spectral measurements were taken in two principal orientations of the electric field **E**-vector of the probing THz and IR radiation relative to the crystallographic *c*-axis, i.e., **E**⊥*c* (perpendicular polarization) and **E**∥*c* (parallel polarization). For that, two plane-parallel slabs were prepared for substituted samples [$x$(Al)=3.0 and $x$(Al)=3.3] with the *c*-axis oriented perpendicular and parallel to the surfaces of the slabs. Due to relatively small crystals size, for the $x$ = 0.0 composition, only the **E**⊥*c* measurement geometry was possible with the *c*-axis oriented perpendicular to the sample's faces. The prepared slabs were carefully polished from both sides utilizing a precision polishing system Allied MultiPrep 8 (with diamond disks with grades down to 1 µm), for the surfaces to meet the requirements of spectral experiments (surface roughness should be much more acceptable compared with the radiation wavelength). Table 2 shows the thicknesses of the prepared samples. A set of THz - IR spectrometers was used to study the spectral response of the compounds. At infrared frequencies, we used Fourier-transform infrared spectrometer Bruker Vertex 80v equipped with a cold finger cryostat to measure the reflectivity spectra $R(v)$ in the frequency range 60 – 8000 cm$^{-1}$. For the measurements at sub-terahertz and terahertz frequencies (8 - 100 cm$^{-1}$), two commercial terahertz time-domain spectrometers, Tera K15 Menlo GmbH and TeraView TPS 3000, equipped with exchange gas cryostats, were used to measure the complex (amplitude and phase) transmission coefficient $T^*(v)$ of the samples from which the spectra of real and imaginary parts of dielectric permittivity were determined, $\varepsilon^*(v, T) = \varepsilon'(v, T) + i\varepsilon''(v, T)$. The THz real and imaginary permittivity spectra were used to calculate the THz reflectivity of the samples that were merged with the measured IR reflectivity spectra. The obtained broad-band THz-IR reflectivities were processed, together with the THz complex permittivity spectra, with the least-square technique to get the broad-band spectra of real and imaginary parts of dielectric permittivity.

Experimental X-ray data sets for $Ba_{0.2}Pb_{0.8}Fe_{12-x}Al_xO_{19}$ compounds were obtained at room temperature on a Bruker D8 QUEST automatic diffractometer (graphite monochromatic Mo-Kα radiation, λ= 0.71073 Å, ω-scan) equipped with PHOTON II detector.

Single crystals of $PbFe_{12}O_{19}$ were grown by a solution-melt crystallization technique by cooling the 39.4PbO ·53.6 $Fe_2O_3$ ·7.0 NaOH mixture from 1350 to 950°C at a speed of 15 grad/hour in the crucibles of yttrium stabilized zirconia. The samples were obtained in the form of hexagonal plates with sizes reaching 10*10*2 mm$^3$. According to the X-ray fluorescence analysis (microanalyzer "EDAX", USA), the cationic chemical composition of the crystals corresponds to the formula $PbFe_{11.4}Zr_{0.6}O_{19}$. The presence of Zr atoms is connected with the interaction of the melt with the crucible material during crystal growth.

The intensity data sets for $PbFe_{11.4}Zr_{0.6}O_{19}$ compound were collected at a temperature of 100(1) K with an Oxford Cryojet at the Advanced Photon Source on Beamline 15ID-D of NSF's ChemMatCARS Sector 15 using Huber 3 circles diffractometer with $\kappa$ angle offset 60° and equipped with Pilatus3X 1 M (CdTe) detector. An unattenuated beam with a wavelength of

0.41328 Å (energy 30 keV) was used with 0.4 s exposure times, and data were collected with $\varphi$ scans at 0.2° increment with $\kappa$ offsets 0.0 and 30° using a 13 cm detector distance. To improve the statistical reliability of the measured data, each frame was measured three times and merged into one frame. Evaluation of the initial X-ray diffraction (XRD) images and reciprocal lattice construction was performed to ascertain the crystal quality. Data images were converted to Bruker format, and integration was performed with APEX II [23] suite software. Data reduction was performed using SAINT v.8.32B software. Scaling and absorption correction was performed by a multi-scan method implemented in SADABS v.2013 program included in the APEX suite. The structural solution and refinements were carried out with SHELX-2014 [24,25] software using the XPREP utility for the space group determination and the XT and XL programs for the structural solution and refinement, respectively. For all of the structures, the positions of metal and oxygen atoms were determined from the Fourier difference map and refined anisotropically, and in each case, the occupancy was determined through free refinement.

## 3. Results and discussion

### Samples characterization

The chemical composition and calculated brutto formulas of grown crystals are summarized in Table 1.

Table 1: Chemical composition (*at.*%) and brutto formula of grown $Ba_{0.2}Pb_{0.8}Al_xFe_{12-x}O_{19}$ crystals.

| Brutto formula | Pb | Ba | Fe | Al | O |
|---|---|---|---|---|---|
| $Ba_{0.2}Pb_{0.8}Fe_{12}O_{19}$ | 2.23 | 0.56 | 33.71 | 0 | 64.62 |
| $Ba_{0.2}Pb_{0.8}Al_3Fe_9O_{19}$ | 2.14 | 0.49 | 23.62 | 7.89 | 65.86 |
| $Ba_{0.2}Pb_{0.8}Al_{3.3}Fe_{8.7}O_{19}$ | 2.1 | 0.51 | 23.07 | 8.74 | 65.56 |

The effect on the lattice parameters of isovalent substitution of iron by aluminum and barium by lead is similar to that found for crystals grown from $Na_2O$ and PbO based fluxes [26–28]. The samples show decreased unit cell parameters compared to pure $BaFe_{12}O_{19}$ that is mainly due to the higher Al-content as compared with the minor Pb-substitution related to the general composition, in conjunction to the larger difference in ionic radii for $Al^{3+}$ and $Fe^{3+}$ (19%) than for $Pb^{2+}$ and $Ba^{2+}$ (7 %): $r(Al^{3+})$ = 39 pm, $r(Fe^{3+})$ = 49 pm for CN = 4; $r(Pb^{2+})$ = 149 pm, $r(Ba^{2+})$ = 161 pm for CN = 12, in a given coordination [29]. Powder X-ray diffraction indicates the presence of single-phase material (Fig. 1). In comparison with a pattern of single crystals of (Ba, Pb)Fe$_{12}$O$_{19}$ obtained from PbO flux without the addition of aluminum oxide [27], the slightly broadened reflections of the Al-substituted samples may indicate some inhomogeneous distribution of Pb and Al within the crystals followed by minor variations in unit cell parameters. The experimental patterns differ from the simulated ones for pure BaFe$_{12}$O$_{19}$ [30] or PbFe$_{12}$O$_{19}$ [31] in peak intensities, which can be explained by texture due to the high aspect ratio of platelet crystals with only minor contribution due to the substitution.

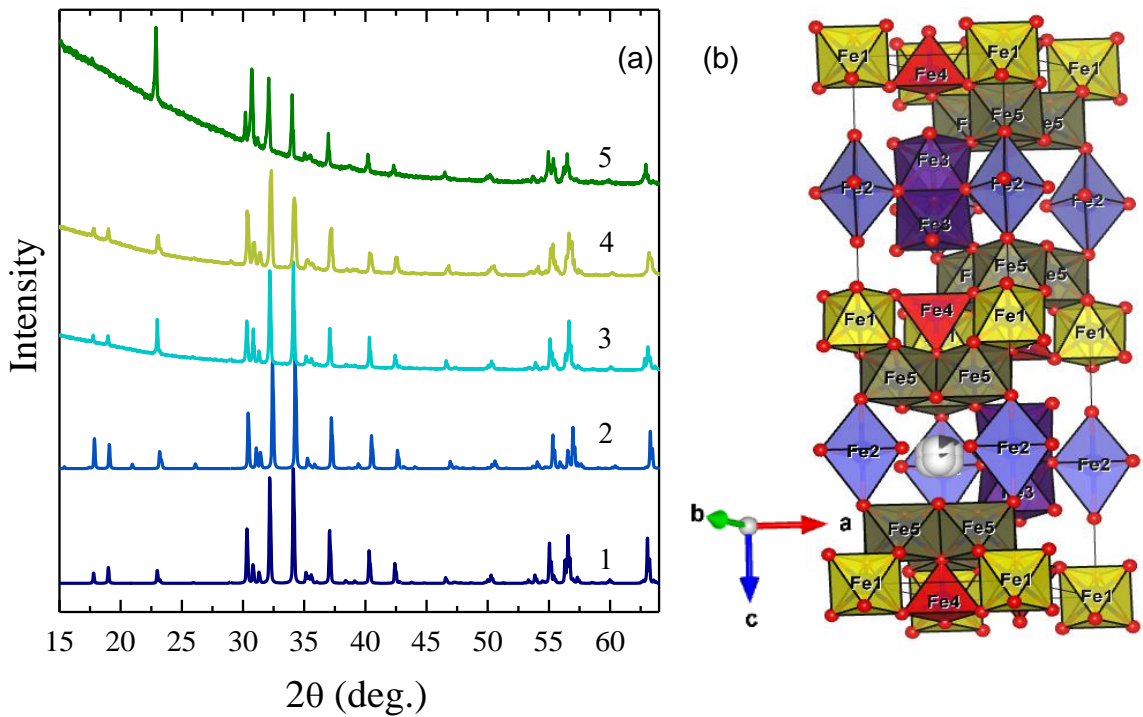

Fig. 1. Panel (a): powder X-ray diffraction patterns of: 1 - BaFe$_{12}$O$_{19}$ (simulated from literature crystal structure data of [49]), 2 - PbFe$_{12}$O$_{19}$ (simulated from literature crystal structure data of [50]), 3 - Ba$_{0.2}$Pb$_{0.8}$Al$_3$Fe$_9$O$_{19}$, 4 - Ba$_{0.2}$Pb$_{0.8}$Al$_{3.3}$Fe$_{8.7}$O$_{19}$, 5 - Ba$_{0.2}$Pb$_{0.8}$Fe$_{12}$O$_{19}$. Panel (b): structure of the M-type hexaferrite: blue cuboctahedra denote 2d site-positions occupied with

Ba/Pb ions, iron site positions are as following: Fe(1) are 2a, Fe(2) - 2b, Fe(3) - $4f_2$, Fe(4) - $4f_1$, and Fe(5) - 12k.

**Terahertz and infrared dielectric properties.**

The room-temperature THz-IR reflectivity spectra of the compounds are shown in Fig. 2. Prominent structures are clearly seen that are caused by absorption lines. These lines were modeled by regular Lorentzian expression for complex dielectric permittivity:

$$\varepsilon^* = \varepsilon' + i\varepsilon'' = \sum_j \frac{f_j}{\nu_j^2 - \nu^2 + i\nu\gamma_j} + \varepsilon_\infty \quad (1).$$

Here $f_j = \Delta\varepsilon_j \nu_j^2$ is the oscillator strength of j-th resonance, $\Delta\varepsilon_j$ is its dielectric strength, $\nu_j$ is the resonance frequency, $\gamma_j$ is the damping factor, and $\varepsilon_\infty$ is the high-frequency dielectric permittivity. The least-square fitting procedure also included simultaneous processing of the THz $\varepsilon'(\nu)$ and $\varepsilon''(\nu)$ spectra as mentioned above. Since all absorption lines were observed at frequencies below 1000 cm$^{-1}$, in the following, we do not present the spectra obtained for higher frequencies. However, the dispersionless reflectivity measured above 1000 cm$^{-1}$ was used to determine the high-frequency dielectric permittivity $\varepsilon_\infty$ that was supposed to be temperature-independent; its values for the studied samples $Ba_{0.2}Pb_{0.8}Al_xFe_{12-x}O_{19}$ with $x$(Al)=0.0, 3.0, and 3.3 are presented in Table 2.

A rich set of absorption lines observed at frequencies above 70 cm$^{-1}$ is attributed to polar lattice phonons [21,22,32–35]. Processing the spectra with Eq.1 allowed to obtain temperature dependence of the parameters ($\Delta\varepsilon_j, \nu_j, \gamma_j,$ and $f_j$) of IR phonon bands for both polarizations, as discussed below. Parameters of all resolved polar phonons of $E_{1u}$ and $A_{2u}$ symmetries are provided in Supplemental Information (SI), Tables SI1-SI6. Already at this stage, one can see that the reflectivity spectra (Fig. 2) are strongly anisotropic and show a distinct shift to higher frequencies of some phonon lines in compounds with higher aluminum content. Authors of [33,36] reported on a soft $A_{2u}$ phonon for **E**||$c$ polarization that was considered as a precursor of a displacive phase transition happening outside of used temperature interval, below $T = 5$ K. We did not detect any signature of such temperature-unstable line in the studied compounds for **E**||$c$ polarization: the lowest $A_{2u}$ phonon in $Ba_{0.2}Pb_{0.8}Al_xFe_{12-x}O_{19}$ is located at the same frequency of ≈75 cm$^{-1}$ as in [33,36]; however, it does not show any significant red-shift. Note that in [33,36], the phonon frequency is shown to decrease down to 35-40 cm$^{-1}$ at helium temperature.

Table 2: High-frequency dielectric permittivity $\varepsilon_\infty$ and thickness $d$ of the studied samples $Ba_{0.2}Pb_{0.8}Al_xFe_{12-x}O_{19}$, $x(Al)=0.0$, 3.0, and 3.3. For $x(Al)=0$, measurements for only $\mathbf{E}\perp c$ polarization were possible due to the small sizes of the grown crystals.

| Concentration | $\mathbf{E}\perp c$ | | $\mathbf{E}||c$ | |
|---|---|---|---|---|
| | $\varepsilon_\infty$ | $d$, μm | $\varepsilon_\infty$ | $d$, μm |
| $x(Al)=0$ | 6.44 | 126 | - | - |
| $x(Al)=3$ | 5.60 | 100 | 4.24 | 100 |
| $x(Al)=3.3$ | 5.50 | 80 | 3.33 | 129 |

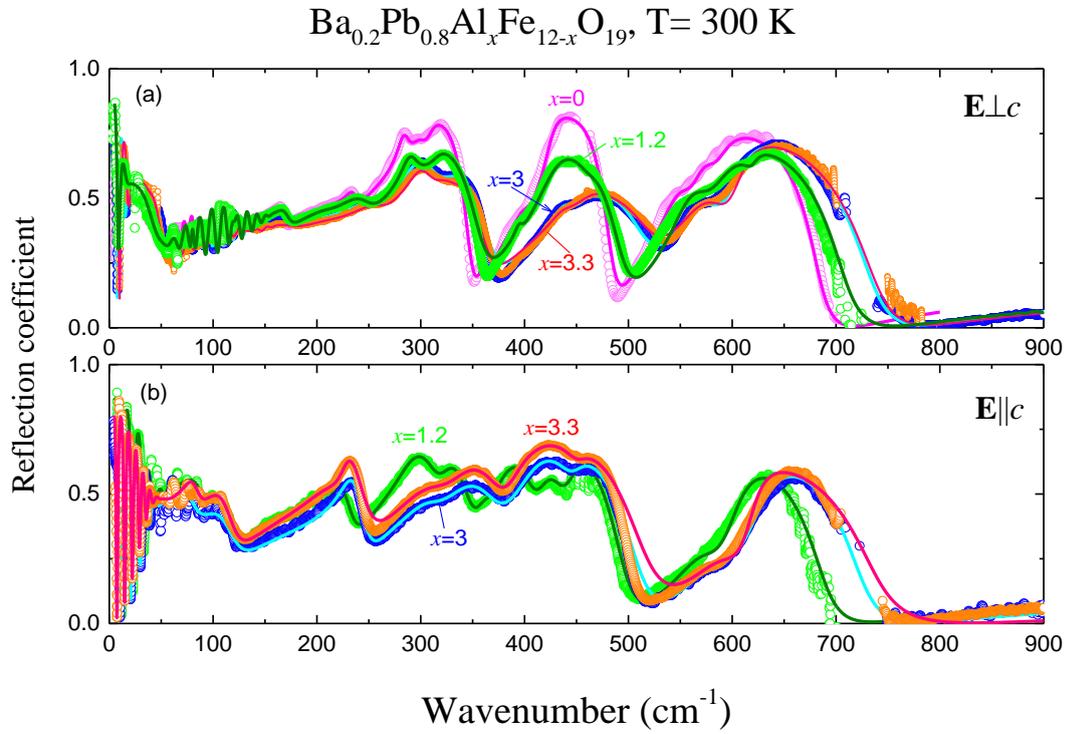

Fig. 2. Reflectivity spectra of single-crystalline $Ba_{0.2}Pb_{0.8}Al_xFe_{12-x}O_{19}$, $x(Al)$=0.0, 1.2, 3.0, 3.3, obtained at $T$ = 300 K for polarizations $\mathbf{E}\perp c$ (a) and $\mathbf{E}\|c$ (b). Dots are the experimental data; solid lines are the results of the least-square fits with Eq. (1) used to describe the absorption resonances and with standard Fresnel expressions [33] to model the spectra of reflectivity of the plane-parallel samples. The oscillations are seen at terahertz frequencies, 60-150 $cm^{-1}$, for $\mathbf{E}\perp c$ and below 50 $cm^{-1}$ for $\mathbf{E}\|c$, manifesting the Fabry-Perot interference within plane-parallel slabs. Data for $x(Al)$=1.2 are taken from Ref. [22].

According to the factor-group analysis [33], BaM has 30 IR-active modes: 17 $E_{1u}$ for $\mathbf{E}\perp c$ polarization and 13 $A_{2u}$ for $\mathbf{E}\|c$ polarization. For the polarization $\mathbf{E}\perp c$, we observe 14 $E_{1u}$ polar phonon modes for $x(Al)$=0, 14 $E_{1u}$ polar phonon modes for $x(Al)$=1.2, 12 $E_{1u}$ polar phonon modes for $x(Al)$=3.0, and 13 $E_{1u}$ polar phonon modes for $x(Al)$= 3.3 (see Fig. 2, panel a). For polarization $\mathbf{E}\|c$, we observe 13 $A_{2u}$ polar phonon modes for $x(Al)$=1.2, 9 $A_{2u}$ polar phonon modes for $x(Al)$=3.0, and 11 $A_{2u}$ polar phonon modes for $x(Al)$=3.3 (see Fig. 2, panel b). We believe that we do not resolve in the obtained spectra a full number of predicted phonon lines because of their broadening and because of their mutual shift when high amounts of Pb and Al are introduced in the crystal lattice. From Fig. 2, it is also seen that doping with aluminum leads to the blue shift of phonon peaks that is in agreement with the results obtained in Ref [37] for $BaAl_xFe_{12-x}O_{19}$ with $0 \leq x(Al) \leq 3.5$. The blue shift is caused by the lower mass of $Al^{3+}$ ions compared with $Fe^{3+}$ ions [29]. The change in the intensity and broadening of the IR phonons accompanying substitution can be assigned to the structural disorder induced by Pb and Al in the crystal lattice; a certain role can also be played by quantum effects [38]. One important notice here is that except for the broadening of the IR bands (Fig. 2), no new phonon peaks were observed in the samples with different Al concentrations confirming the single-phase form of our single crystals.

According to Refs. [37,39], the IR peaks observed at 480-585 $cm^{-1}$ are characterization peaks for Fe-O bonds vibrations at their site positions (octahedral and tetrahedral). The effect of substituting Fe with Al manifests itself in the change of corresponding lattice vibrations in IR spectra (peaks at 480-585 $cm^{-1}$ in Fig. 2); this will be analyzed below in more detail when discussing the THz dielectric response of the compounds.

The temperature and doping evolutions of the IR spectra of the imaginary part of dielectric permittivity measured at two polarizations are shown in Fig. 3. For perpendicular polarization $\mathbf{E}\perp c$, (panels a and b), the compounds without aluminum reveal narrower absorption lines compared with the substituted compounds where the lines are broader (corresponding vibrations are more damped) due to the structural distortion and disorder. Three clear broad peaks are observed at around 280 cm$^{-1}$, 420 cm$^{-1}$, and 550 cm$^{-1}$, and the intensity of these peaks decreases with Al increments; such increments also lead to slight blue shifts of the peaks. For parallel polarization $\mathbf{E}\|c$ (panels c and d), the dielectric response contains resonances that are spread over a wider spectral range than in the case of the perpendicular polarization—for both polarizations, cooling down leads to the noticeable narrowing of some peaks.

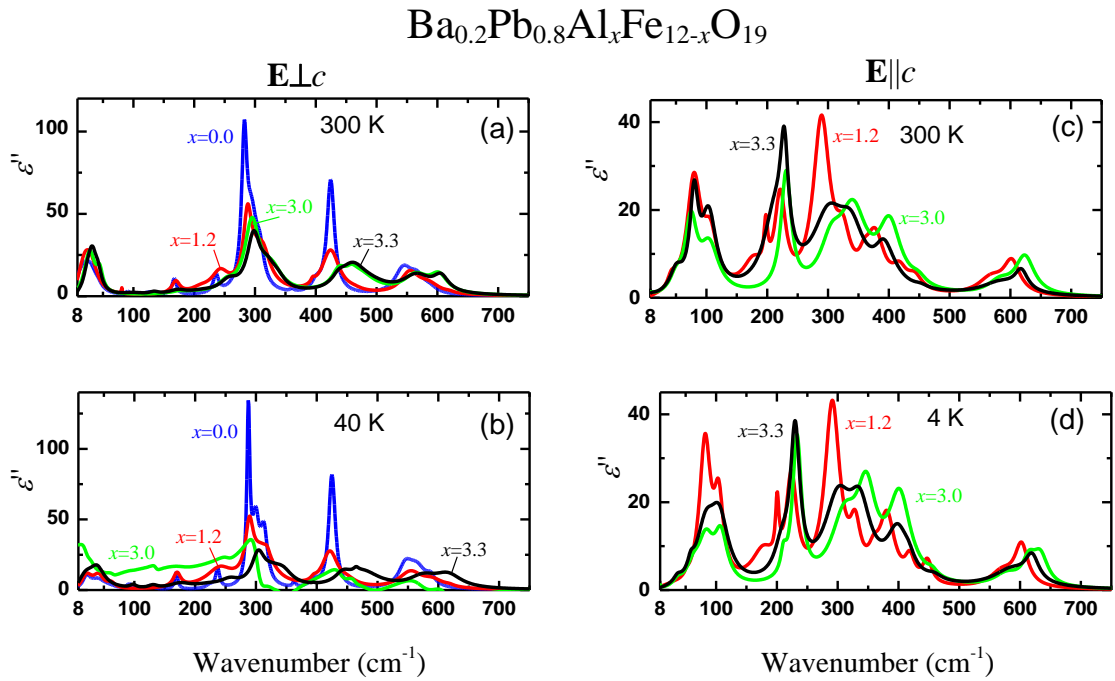

Fig. 3. Infrared spectra of imaginary permittivity of single-crystalline $Ba_{0.2}Pb_{0.8}Al_xFe_{12-x}O_{19}$, $x(Al)=0.0$, 1.2, 3.0, and 3.3 obtained with the least-square fits with Eq. (1) as described in the text. Spectra are shown for polarizations $\mathbf{E}\perp c$ (panels a, b), $\mathbf{E}\|c$ (panels c, d) and for temperatures $T = 300$ K (panels a, c), $T = 40$ K and $T = 4$ K (panels b, d, respectively).

THz spectra of real $\varepsilon'$ and imaginary $\varepsilon''$ parts of dielectric permittivity of the studied compounds measured at various temperatures for two polarizations are presented in Fig. 4 and Fig. 5. Note that the low-frequency (below 10-20 cm$^{-1}$) values of the real permittivity $\varepsilon'$ for perpendicular polarization are larger than the $\varepsilon'$ values for parallel polarization. The same was detected for single-crystalline $BaFe_{12}O_{19}$ [32], where the dielectric permittivity $\varepsilon'$ for $\mathbf{E}\|c$ polarization was about three times smaller than its value for the $\mathbf{E}\perp c$ polarization.

The broad absorption band detected at the THz frequencies for both polarizations comprises peaks connected with the absorption due to electronic transitions within the fine-structured ground state of $Fe^{2+}$ ($^5E$) ions [21]. In figures 4 and 5, these peaks are indicated by arrows and refer to the spectra measured at liquid helium temperature. For the perpendicular polarization, $\mathbf{E}\perp c$, we detect four lines for compounds with $x(Al)=0.0$ and $x(Al)=3.0$ and three lines – for the compound with $x(Al)=3.3$. For parallel polarization, $\mathbf{E}\|c$, three lines are detected for $x(Al)=3.0$ and $x(Al)=3.3$. The fine structure in the spectra is better distinguished for the $\mathbf{E}\perp c$ polarization, the reason being the smaller overlap of electronic peaks with the lowest-frequency IR phonon. For $\mathbf{E}\perp c$, the phonon is located at ~ 90 cm$^{-1}$ and has dielectric strength $\Delta\varepsilon < 0.5$ for $x(Al)=3.0$ and $x(Al)=3.3$, while for $\mathbf{E}\|c$ the overlap is larger since the phonon is located closer to the electronic structure, at a lower frequency of ~75 cm$^{-1}$, and it is more intensive ($\Delta\varepsilon>4$) than for the $\mathbf{E}\perp c$ polarization.

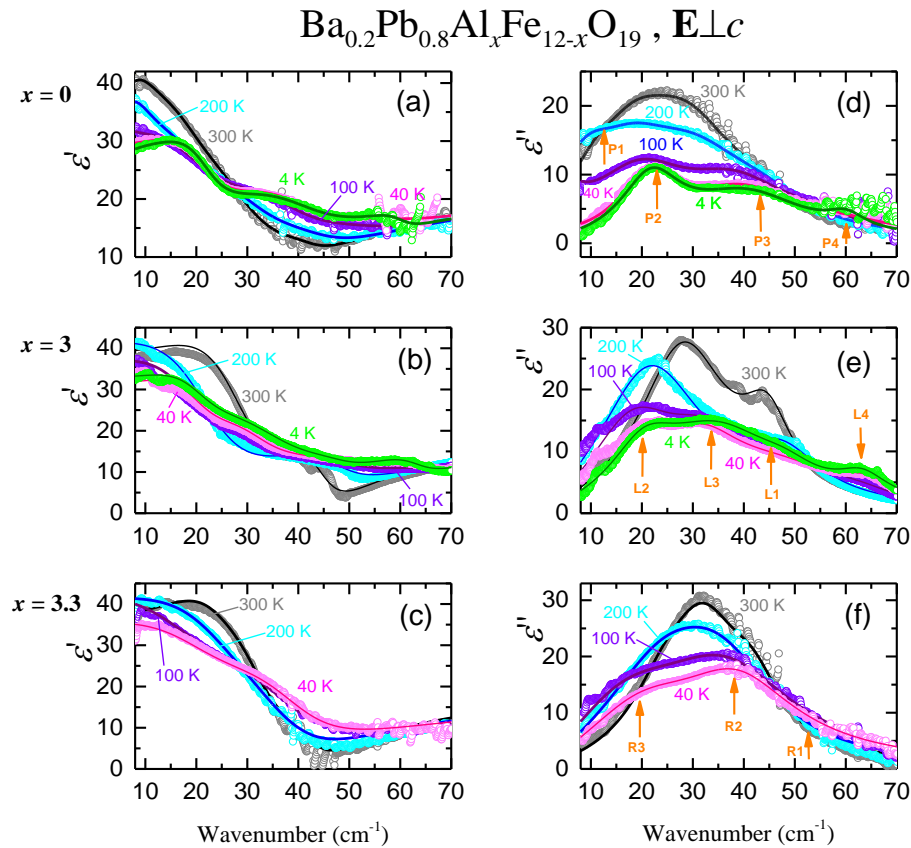

Fig. 4. Temperature-dependent terahertz spectra for polarization $\mathbf{E}\perp c$ of real (panels a, b, c) and imaginary (panels d, e, f) parts of complex dielectric permittivity of single-crystalline $Ba_{0.2}Pb_{0.8}Al_xFe_{12-x}O_{19}$ for $x(Al)=0.0$ (a, d), $x(Al)=3.0$ (b, e), and $x(Al)=3.3$ (c, f). Dots are the experimental data; solid lines are the results of the least-square fits with Eq. (1) used to describe the absorption resonances. Arrows in panels d, e, and f mark frequency positions of electronic transitions within the fine-structured ground state of $Fe^{2+}$ ($^5E$) ions [three for $x(Al)=0.0$ and 3.3, and four for $x(Al)=3.0$].

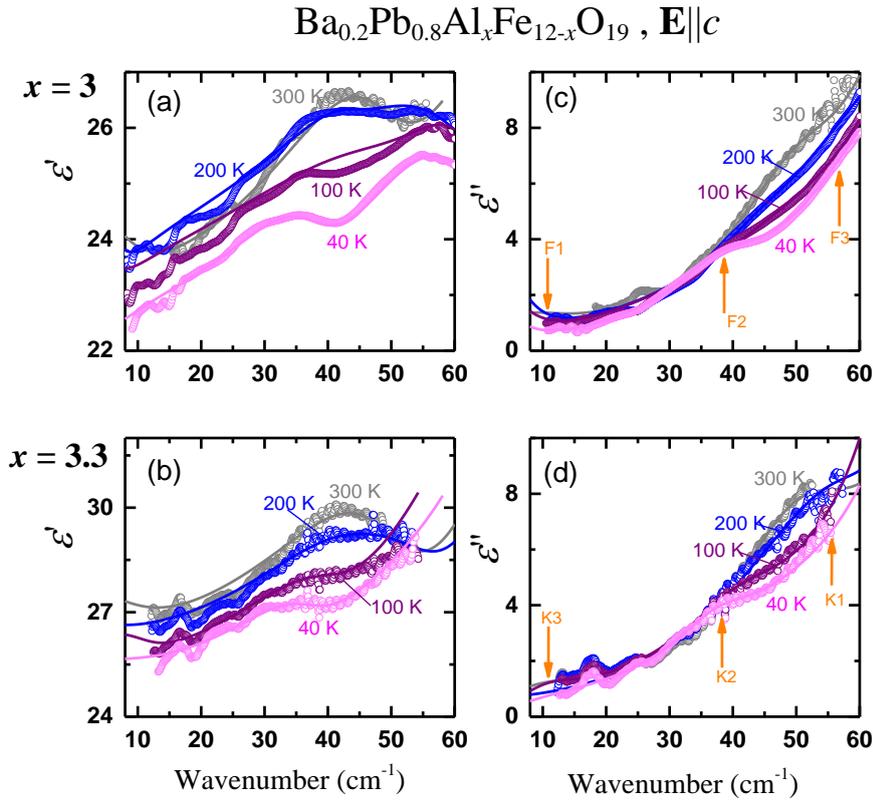

Fig. 5. Temperature-dependent terahertz spectra for polarization $\mathbf{E}\|c$ of real (a,b) and imaginary (c,d) parts of complex dielectric permittivity of single-crystalline $Ba_{0.2}Pb_{0.8}Al_xFe_{12-x}O_{19}$ for $x(Al)=3$ (a, c) and $x(Al)=3.3$ (b,d). Dots are the experimental data; solid lines are the results of the least-square fits with Eq. (1) used to describe the absorption resonances. Arrows in panels c, d mark frequency positions of electronic transitions within the fine-structured ground state of $Fe^{2+}$ ($^5E$) ions.

In Fig. 6 and Fig. 7, we show in more detail the dependence on aluminum content of the THz absorption bands observed for both polarizations and for two selected temperatures, 300 K and 40 K. For $\mathbf{E}\perp c$ (Fig. 6), for both temperatures, the bands in the spectra of imaginary permittivity $\varepsilon''$ (panels b, d) shift towards high frequencies and increase in intensity when the concentration of aluminum increases. Spectra of real permittivity behave correspondingly. The blue shift of the bands can be explained if one considers lattice distortions produced by the introduction of aluminum ions into site-positions of iron ions. The more prominent are the distortions, the larger are local crystal fields, which results in a larger distance between the electronic levels in the fine-structured $^5E$ state of $Fe^{2+}$ ions and, as a result, in a higher frequency position of corresponding peaks in the measured spectra. The growth of the intensity of terahertz absorption with the increase of aluminum content can be associated with the increase of effective concentration of absorption centers, i.e., of $Fe^{2+}$ ions [21]. Since we have a constant concentration of $Pb^{2+}$ ions in the lattice (which is the main source of $Fe^{2+}$ in our system), this increase of $Fe^{2+}$ concentration should be attributed to the effect of aluminum on the site occupancy in the lattice. For the polarization $\mathbf{E}\|c$ (Fig. 7), the evolution of the spectra with aluminum content change is not as clearly seen as in the case of perpendicular polarization due to more substantial overlap of electronic transitions with the IR phonon, as was discussed above. Comparing the spectra of Fig. 6 and Fig. 7, it is obvious that the THz dielectric losses $\varepsilon''$ are significantly lower for the $\mathbf{E}\|c$ polarization than the $\mathbf{E}\perp c$ polarization. Real permittivity values $\varepsilon'$ reveal significant and non-monotonous changes acquiring maximum values for $x(Al)=1.2$. Both observations are most probably connected mainly with the behavior of the lowest-frequency IR phonon. We note that the presented information on the THz dielectric response of the compounds and its sensitivity to doping can be used when designing electronic devices operating at terahertz frequencies [40].

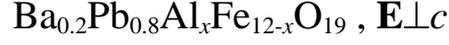

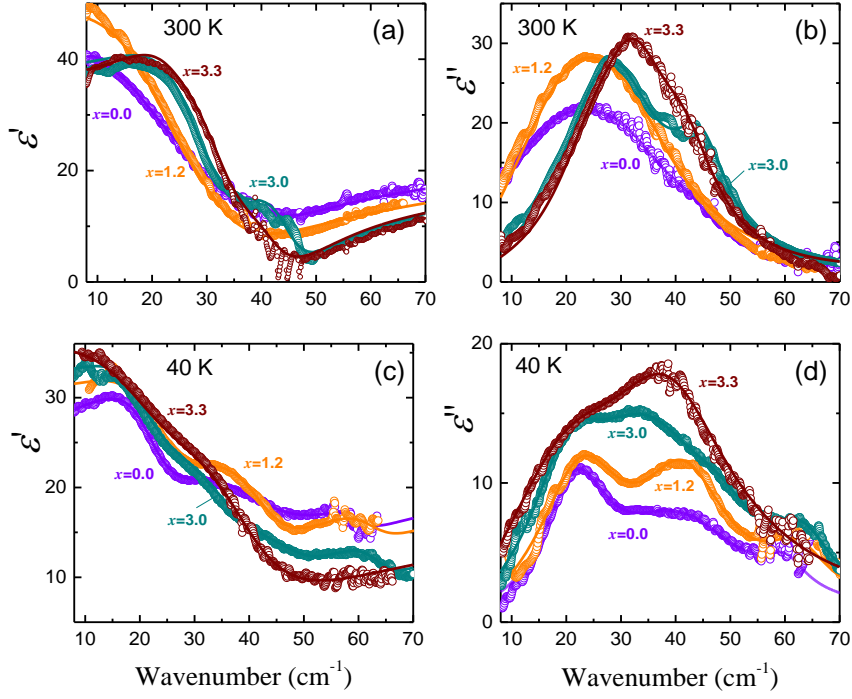

Fig. 6. Terahertz spectra of real (panels a, c) and imaginary (panels b, d) parts of complex dielectric permittivity of single-crystalline $Ba_{0.2}Pb_{0.8}Al_xFe_{12-x}O_{19}$ with $x$(Al)=0.0, 1.2, 3.0 and 3.3 measured for polarization $\mathbf{E}\perp c$ at 300 K and 40 K. Dots are the experimental data, and solid lines are the results of the least-square fits with Eq. (1) used to describe the absorption resonances. Data for $x$(Al)=1.2 are taken from Ref [22].

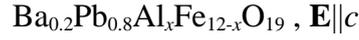

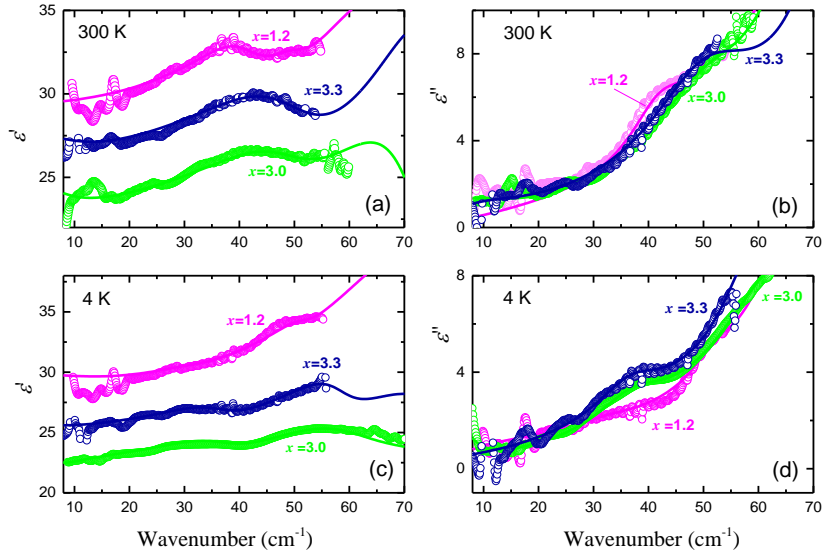

Fig. 7. Terahertz spectra of real (panels a, c) and imaginary (panels b, d) parts of complex dielectric permittivity of single-crystalline $Ba_{0.2}Pb_{0.8}Al_xFe_{12-x}O_{19}$ measured at 300 K and 4 K for polarization $\mathbf{E} \| c$ for $x$(Al)=1.2, 3.0 and 3.3. Dots are the experimental data, and solid lines are the results of the least-square fits with Eq. (1) used to describe the absorption resonances. Data for $x$(Al)=1.2 are taken from ref [22].

**Crystal structure and dielectric response**

In the barium M-type hexaferrite structure, the unit cell is formed by a sequence of S-blocks (cubic close packing) and R-blocks (hexagonal close packing), as indicated in Fig. 1b. Iron ions are distributed over five different site positions: 2a, $4f_2$, 12k are octahedral [Fe (1), Fe (3) and Fe (5)], $4f_1$ is tetrahedral [Fe (4)], and 2b (or 4e) is trigonal bipyramidal [Fe (2)]. Detailed information about the microstructure of BaM is provided in [41]. Though according to the charge neutrality rule, the iron ions in the BaM lattice are expected to be in a trivalent state only, due to several different mechanisms certain amount of divalent iron is also favored in pure BaM, and especially in Pb-substituted BaM [21,22]. According to the results presented in [21,22], the incorporation of lead ions into crystal lattice induces growth of concentration of $Fe^{2+}$ ions. Studies of the Jahn-Teller effect in isostructural compounds [42] show that these $Fe^{2+}$ ions reside in tetrahedral positions solely. According to [21,22] & [43], it is these $Fe^{2+}$ ions that are responsible for the terahertz absorption of the compound, meaning that the larger concentration of $Pb^{2+}$, the more intensive terahertz absorption. All three compounds studied here have the same concentration of lead, i.e., 80%, and thus reveal stronger absorption at THz frequencies than other compounds with a lower concentration of Pb [44].

In Refs. [45] and [27], the authors claim that the preferable residence of the Al ions in the lattice is dependent on the concentration. At low concentration levels, aluminum ions enter preferably octahedral site positions, while during further concentration increases, they start to occupy the tetrahedral site positions as well. According to [40], in $BaFe_{12-x}Al_xO_{19}$, at $x$(Al)=1.1, Al ions occupy octahedral site positions only; at higher doping levels, i.e., $x$(Al)= 3.6, they reside in tetrahedral site positions as well. Moreover, when Al content is even higher, $x$(Al)=4.8, Al ions occupy tetrahedral site positions in (0.82/0.18) ratio (see Table 4 in Ref [27]).

With the above in mind, we are led to the conclusion that in the studied compounds $Ba_{0.2}Pb_{0.8}Fe_{12-x}Al_xO_{19}$ with $x(Al)=3.0$, 3.3 as well as in previously reported compound with $x(Al)=1.2$ [22], aluminum ions practically do not occupy tetrahedral site positions and thus do not interfere with the $Fe^{2+}$ subsystem. To further verify this and to clarify the dynamic behavior of the ions at sublattice sites, a detailed microstructural X-ray study has been performed. Analysis of the obtained data shows that for all compositions, $Al^{3+}$ ions mainly occupy the 2a and 12k octahedral site positions. For $x(Al)=1.2$, occupation degree of 2a positions is 42.8%, and that of 12k positions is 12.4%, while at higher doping levels, $x(Al)=3.0$, 3.3, the respective occupations become close to 83% and 40%. At the same time, occupation of tetrahedral $4f_1$ sites and split bipyramidal 4e sites ($Fe^{2+}$ ions presumably occupy these sites) grows up with an increase of Al content, from 4.8% up to 11.6 % for $4f_1$, and from 1.2% to 6.6% for 4e (see Table 3). Thus, the degree of substitution of iron in tetrahedral positions is not substantial for all studied compounds, in agreement with our conclusions made on the basis of spectroscopic results.

Table 3. Ratio of Al substituted Fe ions for single-crystalline $Ba_{0.2}Pb_{0.8}Al_xFe_{12-x}O_{19}$, ($x(Al)$ = 1.2, 3.0, and 3.3) measured at 300 K using automatic diffractometers Bruker D8 QUEST (graphite monochromatic Mo-Kα radiation, λ = 0.71073 Å, ω scanning).

|         |           | Crystal       |               |               |
|---------|-----------|---------------|---------------|---------------|
| Atoms   | site      | $x(Al)=1.2$   | $x(Al)=3.0$   | $x(Al)=3.3$   |
| Fe1 : Al1 | 2a      | 0.578 : 0.428 (6) | 0.240 : 0.826 (6) | 0.161 : 0.839 (6) |
| Fe2 : Al2 | 4e (2b) | 0.988 : 0.012 (6) | 0.958 : 0.042 (6) | 0.934 : 0.066 (6) |
| Fe3 : Al3 | 4f ($4f_2$) | 0.943 : 0.057 (5) | 0.885 : 0.115 (5) | 0.851 : 0.149 (5) |
| Fe4 : Al4 | 4f ($4f_1$) | 0.952 : 0.048 (4) | 0.919 : 0.081 (5) | 0.884 : 0.116 (5) |
| Fe5 : Al5 | 12k     | 0.876 : 0.124 (3) | 0.660 : 0.340 (3) | 0.602 : 0.398 (3) |

Close inspection of the structural results revealed a remarkable flat displacement ellipsoid for the mixed occupied Pb/Ba site. Such enlarged displacement parameters for Pb were earlier described for pure $PbFe_{12}O_{19}$ [31]. Results of structural refinement of the Pb-Al substituted BaM and PbM are presented in SI, Table SI7; for atomic coordinates, see SI Table SI8.

For all measured $Ba_{0.2}Pb_{0.8}Al_xFe_{12-x}O_{19}$ crystals, Ba (Pb) cations exhibit disorder in their displacements from fixed 2d position (2/3, 1/3, 1/4) to 12j site (x, y, 1/4). The shifts along *x* and *y* directions are varied around 0.25 - 0.3 Å. Pb/Ba ions are equally distributed over six split positions resulting in an occupancy of 0.167. When the lead concentration is raised up to 100%, i.e., in the case of PbM crystal, an additional disorder along the *z*-axis was also observed. Presumably, this disorder can be connected with the influence of the active lone-pair effect of $6s^2Pb^{2+}$ cations. For pure Ba hexaferrite, this kind of disorder is not observing owing to $Ba^{2+}$ cation that has no remaining valence electron. Concentration dependences of the bonds lengths of the studied crystals are provided in SI Table SI9.

**Crystal field distortion**

In Figures 8, 9, and 10, we present detailed temperature dependences of parameters of THz absorption lines that are associated with electronic transitions, as discussed above. We observe 4 lines (polarization $\mathbf{E}\perp c$) in the $x$(Al)=0 compound, and 3 lines in compounds with $x$(Al)=3.0 and $x$(Al)=3.3 (for both polarizations). Interestingly, in the compound with $x$(Al)=3.0, there are two lines, L3 and L4, in the $\mathbf{E}\perp c$ spectra of Fig. 9, that are resolved below 190 K and 110 K, respectively. Since we did not detect any structural change in the X-ray experiments that could lead to activation in the spectra of folding phonon resonances, we associate this observation with the low intensity of these lines at higher temperatures. In all studied compounds, for both polarizations and at all temperatures, the THz lines reveal only a slight change of their positions with temperature; some are blue-shifted, and some are red-shifted. Certain lines, however, show considerable changes in their dielectric ($\Delta\varepsilon$) and oscillator ($f$) strengths and damping factors ($\gamma$). The crystal field distortions play a vital role in our systems, compared with pure BaM or PbM. In our previous work [22], we proposed a model that fully accounts for the origin of the observed THz absorption resonances ascribing them to transitions between energy levels of fine-structure components of $Fe^{2+}$. The model is based on consideration of the trigonal crystal field distortions that lead to lowering the symmetry of the site-positions of iron ions. The model can also be applied to explain the origin and the temperature-doping behavior of the THz electronic resonances observed in the present compounds $Ba_{0.2}Pb_{0.8}Al_xFe_{12-x}O_{19}$, $x$(Al)=0.0, 3.0, and 3.3. Corresponding information is summarized in Table 4. The $A_1 \rightarrow A_2$ transition is magnetic-dipole active for $\mathbf{E}\|c$; the $A_1 \rightarrow E$ transition manifests itself for polarization $\mathbf{E}\perp c$ (active in electric-dipole and magnetic-dipole approximations) and for $\mathbf{E}\|c$ (as electric-quadrupole), the $A_1 \rightarrow A_1$ transition manifest itself for $\mathbf{E}\perp c$ (as electric-quadrupole active) and for $\mathbf{E}\|c$ (as electric-dipole active). According to the symmetry selection rules of the $C_{3v}$ point group, the $A_1 \rightarrow A_2$ transition is magnetic-dipole active and is observed for $\mathbf{E}\|c$ at a frequency $\nu \approx 8 - 15$ cm$^{-1}$[46]. In a compound with zero Al content, we observe a signature of a certain absorption line for polarization $\mathbf{E}\perp c$ below our lowest experimentally available frequency; this was seen as an upturn on the low-frequency wing in the spectra of real and imaginary permittivity. At this stage, we do not have enough information to associate the low-frequency excitation either with overdamped relaxation of the type seen in $Ti^{4+}$-substituted barium hexaferrite [35] or with a ferroelectric-like soft mode detected in Pb-substituted hexaferrites [36,43]. Additional microwave experiments are needed here.

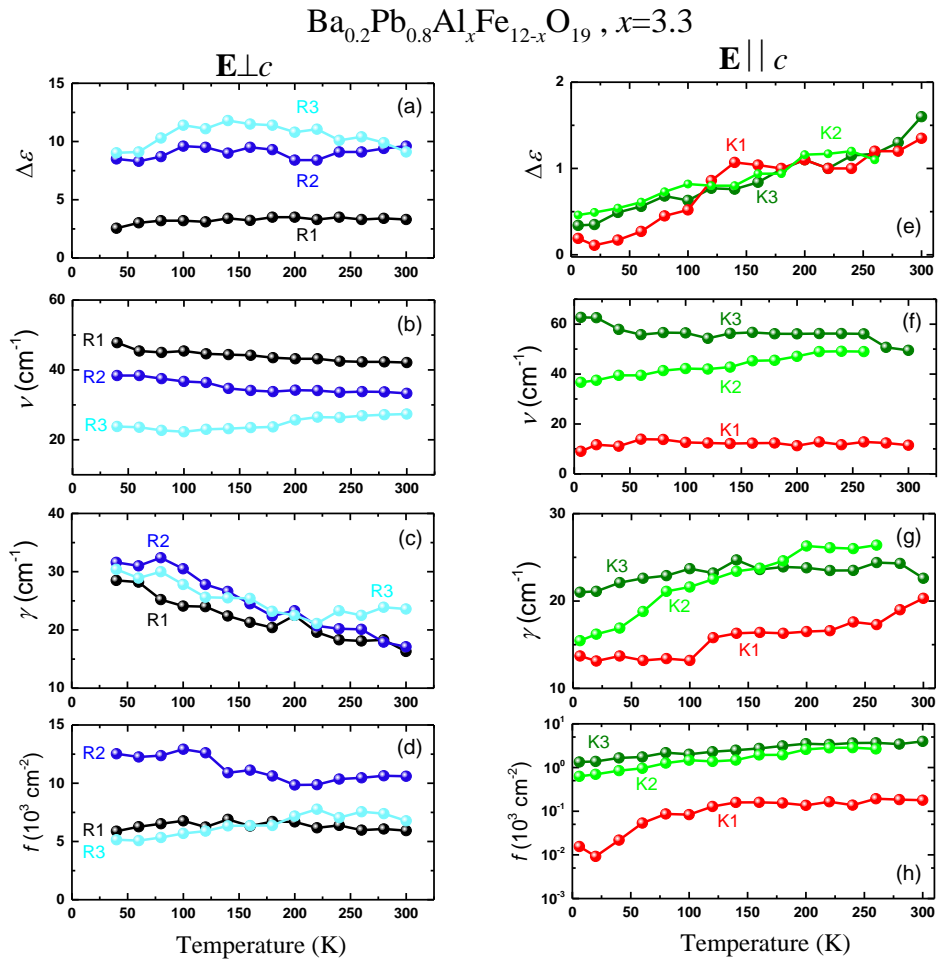

Fig. 8. Temperature dependence of the parameters of the terahertz absorption lines observed for polarizations $\mathbf{E}\perp c$ and $\mathbf{E}\|c$ in single-crystalline $Ba_{0.2}Pb_{0.8}Al_xFe_{12-x}O_{19}$, $x(Al)=3.3$: dielectric strength $\Delta\varepsilon$ (a,e), resonance frequency $\nu$ (b, f), damping factor $\gamma$ (c, g), and oscillator strength $f$ (d, h).

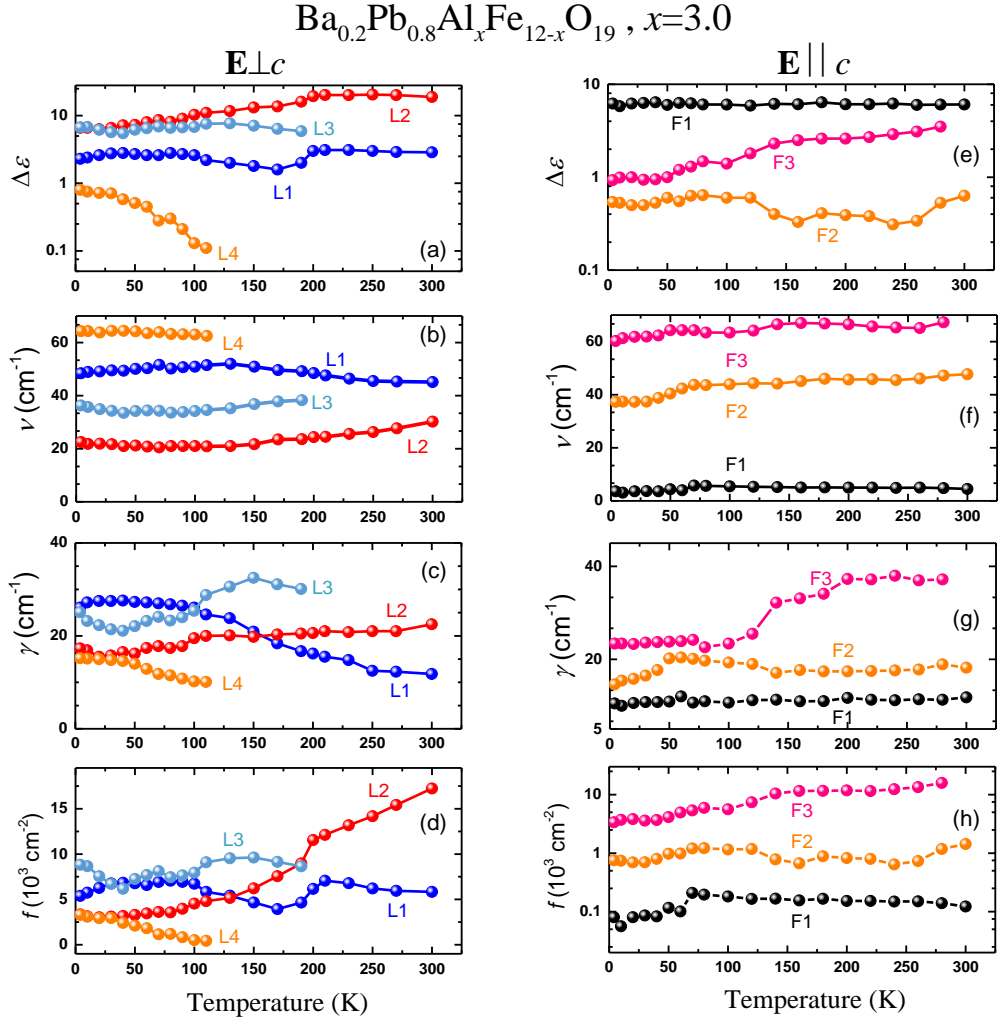

Fig. 9. Temperature dependence of the parameters of the terahertz absorption lines observed for polarizations $\mathbf{E}\perp c$ and $\mathbf{E}\|c$ in single-crystalline $Ba_{0.2}Pb_{0.8}Al_xFe_{12-x}O_{19}$, $x(Al)=3.0$: dielectric strength $\Delta\varepsilon$ (a, e), resonance frequency $\nu$ (b, f), damping factor $\gamma$ (c, g), and oscillator strength $f$ (d, h).

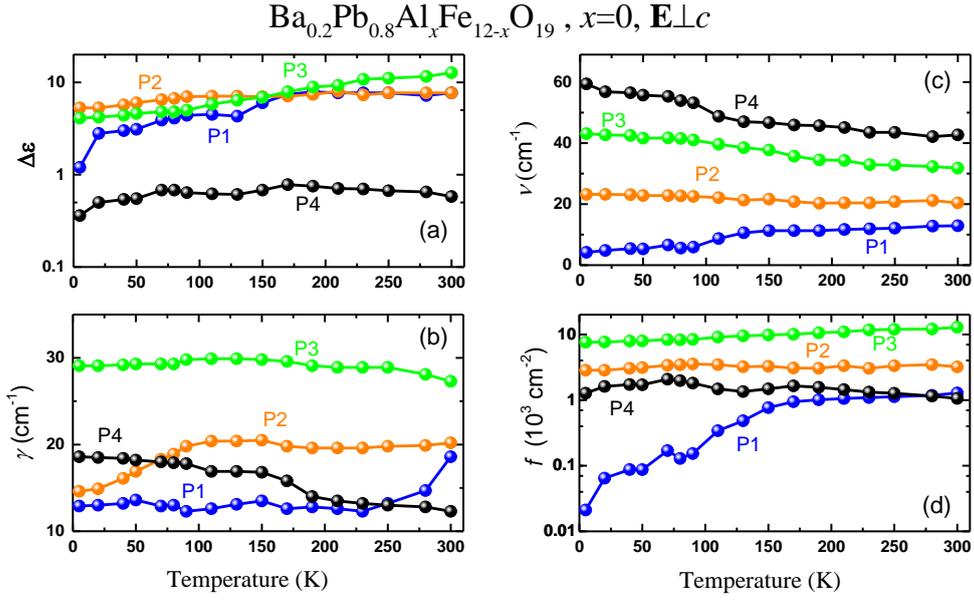

Fig. 10. Temperature dependence of the parameters of the terahertz absorption lines observed for polarization $\mathbf{E}\perp c$ in single-crystalline $Ba_{0.2}Pb_{0.8}Al_xFe_{12-x}O_{19}$, $x(Al)=0$: dielectric strength $\Delta\varepsilon$ (a), resonance frequency $\nu$ (b), damping factor $\gamma$ (c), and oscillator strength $f$ (d).

Table 4: Room-temperature (and lower-temperature as shown in brackets) frequency positions (in $cm^{-1}$) of transitions within the fine structure components of the ground state $^5E$ of divalent iron in single-crystalline $Ba_{0.2}Pb_{0.8}Al_xFe_{12-x}O_{19}$ samples with $x(Al) = 0.0, 3.0, 3.3$ for polarizations $\mathbf{E}\|c$ and $\mathbf{E}\perp c$.

|  | x(Al)=3.3 | | | x(Al)=3.0 | | | x(Al)=0.0 | | |
|---|---|---|---|---|---|---|---|---|---|
|  | Line | Polarization | $\nu$ (cm$^{-1}$) | Line | Polarization | $\nu$ (cm$^{-1}$) | Line | Polarization | $\nu$ (cm$^{-1}$) |
| $A_1$-$A_2$ ($T_1$) | K1 | $\mathbf{E}\|c$ | 11 | F1 | $\mathbf{E}\|c$ | 4 |  |  |  |
| $A_1$-E ($T_1$) | R3 | $\mathbf{E}\perp c$ | 25 | L2 | $\mathbf{E}\perp c$ | 30 | P2 | $\mathbf{E}\perp c$ | 21 |
| $A_1$-E (E) | R2 | $\mathbf{E}\perp c$ | 35 | L3 | $\mathbf{E}\perp c$ | 35 (at 190 K) | P3 | $\mathbf{E}\perp c$ | 31 |
| $A_1$-$A_1$ ($T_2$) | K2 | $\mathbf{E}\|c$ | 48 (at 260 K) | F2 | $\mathbf{E}\|c$ | 44 |  |  |  |
| $A_1$-E ($T_2$) | R1 | $\mathbf{E}\perp c$ | 42 | L1 | $\mathbf{E}\perp c$ | 48 | P4 | $\mathbf{E}\perp c$ | 42 |
| $A_1$-$A_2$($A_2$) | K3 | $\mathbf{E}\|c$ | 56 | F3 | $\mathbf{E}\|c$ | 65 |  |  |  |

| | | | (at 260K) | | | (at 280 K) | | | |
| | | | | L4 | $\mathbf{E} \perp c$ | 63 | | | |
| | | | | | | (at 110 K) | | | |

## 4. Conclusions

The electrodynamic response of single-crystalline $Ba_{0.2}Pb_{0.8}Al_xFe_{12-x}O_{19}$ compounds with $x$(Al)=0.0, 3.0, 3.3, prepared by flux techniques, was investigated at frequencies 8 - 8000 cm$^{-1}$ in the temperature interval 4 – 300 K and for two principal orientations of electric field vector relative to the $c$-axis, $\mathbf{E} \| c$ and $\mathbf{E} \perp c$. The dielectric properties are shown to be strongly anisotropic and strongly dependent on the concentration of aluminum ions. The origin of discovered terahertz absorption resoinances is assigned to electronic transitions between non-degenerated and double-degenerated sub-levels of the $Fe^{2+}$ ground state. In contrast to the data reported for pure $BaFe_{12}O_{19}$ and for compounds with low (up to 30%) lead content, no sign of the softening of the lowest $A_{2u}$ polar phonon is detected in $Ba_{0.2}Pb_{0.8}Al_xFe_{12-x}O_{19}$, indicating suppression of possible displacive phase transition. Analysis of dielectric and X-ray data allows concluding that for all concentrations of $Al^{3+}$ ions, $x$(Al)=1.2, 3.0, 3.3, they mainly occupy the $2a$ and $12k$ octahedral site positions and that the degree of substitution of iron in tetrahedral positions is not substantial. We demonstrate that terahertz-infrared spectroscopies provide detailed quantitative data on the dielectric properties of the compounds, on the dynamical behavior of the atoms and their site positions, the information that is vital for the development of next-generation electronic devices.

**Acknowledgments**

At MIPT, the terahertz study was supported by the Russian Science Foundation (grant 21-79-10184), infrared research was supported by RFBR (grant 20-32-90034). The single crystal growth was supported at SUSU by RFBR (grant 20-38-70057). X-ray structural studies were supported by RFBR (grant 20-03-00337). A. Ahmed is partially funded by a scholarship from the Ministry of Higher Education of the Arab Republic of Egypt. A.S. and A.B. acknowledge-the support of the Ministry of Science and Higher Education of the Russian Federation. L.A. acknowledges system of scholarships of the President of the Russian Federation for young researchers (SP-777.2021.5). NSF's ChemMatCARS Sector 15 is supported by the Divisions of Chemistry (CHE) and Materials Research (DMR), National Science Foundation, under grant number NSF/CHE- 1834750. Use of the Advanced Photon Source, an Office of Science User Facility operated for the U.S. Department of Energy (DOE) Office of Science by Argonne National Laboratory, was supported by the U.S. DOE under Contract No. DE-AC02-06CH11357.

# Supplementary Information

**Infrared polar phonons**

A rich set of absorption lines was observed in the spectra of reflection coefficient at frequencies 70-1000 cm$^{-1}$ and assigned to polar lattice phonons [1–6]. Processing the spectra with the model of free Lorentzian (Eq. 1 in the main text) allowed to obtain temperature-dependent parameters (dielectric strengths $\Delta\varepsilon_j$, resonance frequencies $\nu_j$, damping factors $\gamma_j$ and oscillator strengths $f_j = \Delta\varepsilon_j \nu_j^2$) of infrared phonon bands for both studied polarizations. In tables SI (1-6), we collect parameters of infrared phonons for single-crystalline Ba$_{0.2}$Pb$_{0.8}$Al$_x$Fe$_{12-x}$O$_{19}$, $x$ = 0.0, 3.0, and 3.3, obtained at room temperature and at low temperatures 4 K and 40 K. The obtained spectra resolved with temperature, some absorption lines are resolved only with cooling as indicated in tables SI (1, 3 and 5).

Table **SI**1. Parameters of phonon E$_{1u}$ modes (**E**⊥$c$) for single-crystalline Ba$_{0.2}$Pb$_{0.8}$Al$_x$Fe$_{12-x}$O$_{19}$, $x$(Al) = 0.0: $\Delta\varepsilon$ - dielectric contribution, $\nu$ - frequency, $\gamma$ - damping, $f$ - the oscillator strength, and $\varepsilon_\infty$ - high-frequency permittivity. $T$ =300 K.

| E$_{1u}$ modes (**E**⊥$c$) | $\Delta\varepsilon$ | $\nu$ (cm$^{-1}$) | $\gamma$ (cm$^{-1}$) | $f$ (cm$^{-2}$) |
|---|---|---|---|---|
| 1 | 0.17 | 586.7 | 29 | 58517 |
| 2 | 0.39 | 566.2 | 27 | 125027 |
| 3 | 0.99 | 544.4 | 34 | 293408 |
| 4 | 0.07 | 442.5 | 36 | 13706 |
| 5 | 2.39 | 424.5 | 15 | 430679 |
| 6 | Not resolved at 300 K | | | |
| 7 | 0.26 | 400.6 | 17 | 41725 |
| 8 | 0.03 | 360.8 | 8 | 3905 |
| 9 | 0.89 | 308.1 | 16 | 84484 |
| 10 | 3.35 | 294.4 | 23 | 290349 |
| 11 | 3.04 | 282.5 | 10 | 242611 |
| 12 | 0.48 | 235.8 | 10 | 26689 |
| 13 | Not resolved at 300 K | | | |
| 14 | 0.52 | 167.3 | 9 | 14554 |
| 15 | 0.44 | 129.2 | 28 | 7345 |
| 16 | 0.15 | 90.9 | 7 | 1239 |
| $\varepsilon_\infty$ = 6.44 | | | | |

Table **SI**2. Parameters of E$_{1u}$ modes (**E**⊥$c$) for single-crystalline Ba$_{0.2}$Pb$_{0.8}$Al$_x$Fe$_{12-x}$O$_{19}$, $x$(Al) = 0.0. $\Delta\varepsilon$ - dielectric contribution, $\nu$ - frequency, $\gamma$ - damping, $\varepsilon_\infty$ - high-frequency permittivity. $T$ =4 K.

| E$_{1u}$ modes (**E**⊥$c$) | $\Delta\varepsilon$ | $\nu$ (cm$^{-1}$) | $\gamma$ (cm$^{-1}$) | $f$ (cm$^{-2}$) |
|---|---|---|---|---|
| 1 | 0.14 | 585.5 | 16 | 47993 |
| 2 | 0.77 | 566.6 | 36 | 247197 |
| 3 | 0.74 | 547.8 | 28 | 222063 |
| 4 | 0.03 | 436.5 | 12 | 5716 |
| 5 | 1.30 | 425.4 | 11 | 235255 |
| 6 | 0.99 | 421.1 | 16 | 175552 |
| 7 | 0.14 | 394.6 | 11 | 21799 |
| 8 | 0.08 | 363.8 | 7 | 10588 |
| 9 | 1.76 | 314.1 | 16 | 173640 |
| 10 | 1.54 | 301.3 | 14 | 139804 |
| 11 | 3.57 | 287.6 | 8 | 295288 |
| 12 | 0.31 | 237.5 | 7 | 17486 |
| 13 | 0.12 | 221.7 | 11 | 5898 |
| 14 | 0.45 | 169.5 | 6 | 12929 |
| 15 | 0.57 | 123.7 | 38 | 8722 |
| 16 | 0.16 | 88.7 | 18 | 1259 |
| $\varepsilon_\infty = 6.44$ | | | | |

Table SI3. Parameters of phonon $E_{1u}$ modes ($\mathbf{E} \perp c$) and $A_{2u}$ modes ($\mathbf{E} \| c$) for single-crystalline $Ba_{0.2}Pb_{0.8}Al_xFe_{12-x}O_{19}$, $x(Al) = 3.0$. $\Delta\varepsilon$ - dielectric contribution, $\nu$ - frequency, $\gamma$ - damping, $\varepsilon_\infty$ - high-frequency permittivity. $T = 300$ K.

| $E_{1u}$ modes ($\mathbf{E} \perp c$) | $\Delta\varepsilon$ | $\nu$ (cm$^{-1}$) | $\gamma$ (cm$^{-1}$) | $f$ (cm$^{-2}$) | $A_{2u}$ modes ($\mathbf{E} \| c$) | $\Delta\varepsilon$ | $\nu$ (cm$^{-1}$) | $\gamma$ (cm$^{-1}$) | $f$ (cm$^{-2}$) |
|---|---|---|---|---|---|---|---|---|---|
| 1 | 0.60 | 602.6 | 35 | 217876 | 1 | 0.54 | 624.0 | 39 | 210263 |
| 2 | 0.96 | 564.9 | 50 | 306348 | 2 | Not resolved at 300 K | | | |
| 3 | 1.71 | 471.6 | 66 | 380315 | 3 | 0.22 | 583.1 | 47 | 74801 |
| 4 | 0.19 | 458.8 | 21 | 39995 | 4 | 0.13 | 448.8 | 31 | 26185 |
| 5 | 0.61 | 438.2 | 28 | 117132 | 5 | 1.68 | 401.8 | 46 | 271225 |
| 6 | 0.12 | 333.8 | 28 | 13371 | 6 | 2.83 | 342.1 | 55 | 331202 |
| 7 | 1.75 | 326.1 | 44 | 186097 | 7 | 1.31 | 306.0 | 44 | 122663 |
| 8 | 4.06 | 295.3 | 29 | 354041 | 8 | 2.49 | 229.8 | 21 | 131492 |
| 9 | 0.63 | 253.8 | 31 | 40581 | 9 | Not resolved at 300 K | | | |
| 10 | 0.87 | 189.7 | 65 | 31308 | 10 | 4.08 | 106.2 | 32 | 46016 |
| 11 | 0.07 | 174.2 | 17 | 2124 | 11 | 5.63 | 72.4 | 31 | 29511 |
| 12 | 0.09 | 91.6 | 34 | 755 | | | | | |
| $\varepsilon_\infty = 5.63$ | | | | | $\varepsilon_\infty = 4.24$ | | | | |

Table SI4. Parameters of $E_{1u}$ modes ($\mathbf{E} \perp c$) and $A_{2u}$ modes ($\mathbf{E} \| c$) for single-crystalline $Ba_{0.2}Pb_{0.8}Al_xFe_{12-x}O_{19}$, $x(Al) = 3.0$. $\Delta\varepsilon$ - dielectric contribution, $\nu$ - frequency, $\gamma$ - damping, $\varepsilon_\infty$ - high-frequency permittivity. $T = 4$ K.

| $E_{1u}$ modes ($\mathbf{E} \perp c$) | $\Delta\varepsilon$ | $\nu$ (cm$^{-1}$) | $\gamma$ (cm$^{-1}$) | $f$ (cm$^{-2}$) | $A_{2u}$ modes ($\mathbf{E} \| c$) | $\Delta\varepsilon$ | $\nu$ (cm$^{-1}$) | $\gamma$ (cm$^{-1}$) | $f$ (cm$^{-2}$) |
|---|---|---|---|---|---|---|---|---|---|
| 1 | 0.62 | 611.1 | 33 | 231535 | 1 | 0.33 | 633.4 | 31 | 132395 |
| 2 | 0.01 | 629.9 | 15 | 3968 | 2 | 0.20 | 614.6 | 25 | 75547 |
| 3 | 1.20 | 573.5 | 57 | 394683 | 3 | 0.25 | 578.6 | 52 | 83694 |
| 4 | 1.70 | 475.0 | 65 | 383563 | 4 | 0.06 | 455.4 | 21 | 12443 |
| 5 | 0.11 | 461.3 | 16 | 23408 | 5 | 2.17 | 402.3 | 45 | 351204 |
| 6 | 0.58 | 438.2 | 28 | 111371 | 6 | 2.17 | 347.5 | 39 | 262041 |
| 7 | 0.64 | 342.4 | 28 | 75032 | 7 | 2.31 | 311.6 | 52 | 224288 |
| 8 | 1.50 | 327.8 | 38 | 161179 | 8 | 2.53 | 233.2 | 18 | 137587 |
| 9 | 3.22 | 301.8 | 24 | 293288 | 9 | 0.24 | 210.9 | 11 | 10675 |
| 10 | 0.19 | 255.9 | 25 | 12442 | 10 | 2.73 | 109.5 | 25 | 32733 |
| 11 | 0.55 | 177.6 | 43 | 17348 | 11 | 4.14 | 80.6 | 32 | 26895 |
| 12 | 0.05 | 96.9 | 10 | 469 | | | | | |
| $\varepsilon_\infty = 5.62$ | | | | | $\varepsilon_\infty = 4.24$ | | | | |

Table **SI5**. Parameters of $E_{1u}$ modes ($\mathbf{E}\perp c$) and $A_{2u}$ modes ($\mathbf{E}\|c$) for single-crystalline $Ba_{0.2}Pb_{0.8}Al_xFe_{12-x}O_{19}$, $x(Al) = 3.3$. $\Delta\varepsilon$ - dielectric contribution, $\nu$ - frequency, $\gamma$ - damping, $\varepsilon_\infty$ - high-frequency permittivity. $T = 300$ K.

| $E_{1u}$ modes ($\mathbf{E}\perp c$) | $\Delta\varepsilon$ | $\nu$ (cm$^{-1}$) | $\gamma$ (cm$^{-1}$) | $f$ (cm$^{-2}$) | $A_{2u}$ modes ($\mathbf{E}\|c$) | $\Delta\varepsilon$ | $\nu$ (cm$^{-1}$) | $\gamma$ (cm$^{-1}$) | $f$ (cm$^{-2}$) |
|---|---|---|---|---|---|---|---|---|---|
| 1 | 0.46 | 606.0 | 30.49 | 168929 | 1 | 0.19 | 617.9 | 26 | 72542 |
| 2 | 1.11 | 566.6 | 57.83 | 356350 | 2 | 0.41 | 585.5 | 80 | 140552 |
| 3 | 0.54 | 493.0 | 68.63 | 131247 | 3 | 0.04 | 452.8 | 29 | 8201 |
| 4 | 1.90 | 461.3 | 54.51 | 404316 | 4 | 0.23 | 404.9 | 51 | 37707 |
| 5 | 0.27 | 437.4 | 25.12 | 51656 | 5 | 0.71 | 392.0 | 37 | 109101 |
| 6 | Not resolved at 300 K | | | | 6 | 1.75 | 338.9 | 52 | 200993 |
| 7 | 0.57 | 337.2 | 31.91 | 64811 | 7 | 3.40 | 302.1 | 65 | 310299 |
| 8 | 1.32 | 320.1 | 33.76 | 135253 | 8 | 1.73 | 228.1 | 16 | 90011 |
| 9 | 2.80 | 297.9 | 25.87 | 248484 | 9 | 3.38 | 213.6 | 45 | 154212 |
| 10 | 0.96 | 257.6 | 36.43 | 63703 | 10 | 5.12 | 104.9 | 29 | 56341 |
| 11 | 0.45 | 216.6 | 59.18 | 21112 | 11 | 4.86 | 78.2 | 24 | 29720 |
| 12 | Not resolved at 300 K | | | | | | | | |
| 13 | 0.47 | 173.3 | 33.57 | 14115 | | | | | |
| 14 | 0.19 | 134.4 | 18.16 | 3432 | | | | | |
| 15 | 0.56 | 101.9 | 46.97 | 5815 | | | | | |
| $\varepsilon_\infty = 5.50$ | | | | | $\varepsilon_\infty = 3.33$ | | | | |

Table **SI6**. Parameters of $E_{1u}$ modes ($\mathbf{E}\perp c$) and $A_{2u}$ modes ($\mathbf{E}\|c$) for single-crystalline $Ba_{0.2}Pb_{0.8}Al_xFe_{12-x}O_{19}$, $x(Al) = 3.3$. $\Delta\varepsilon$ - dielectric contribution, $\nu$ - frequency, $\gamma$ - damping, $\varepsilon_\infty$ - high-frequency permittivity. $T = 40$ K.

| $E_{1u}$ modes ($\mathbf{E}\perp c$) | $\Delta\varepsilon$ | $\nu$ (cm$^{-1}$) | $\gamma$ (cm$^{-1}$) | $f$ (cm$^{-2}$) | $A_{2u}$ modes ($\mathbf{E}\|c$) | $\Delta\varepsilon$ | $\nu$ (cm$^{-1}$) | $\gamma$ (cm$^{-1}$) | $f$ (cm$^{-2}$) |
|---|---|---|---|---|---|---|---|---|---|

| 1  | 0.83 | 616.3 | 55 | 315255 | 1  | 0.31 | 624.8 | 27 | 121016 |
| 2  | 0.71 | 575.2 | 55 | 234907 | 2  | 0.70 | 595.7 | 98 | 248401 |
| 3  | 1.69 | 476.8 | 70 | 384202 | 3  | 0.14 | 458.8 | 33 | 29470  |
| 4  | 0.05 | 464.8 | 10 | 10802  | 4  | 0.30 | 418.6 | 30 | 52568  |
| 5  | 0.47 | 443.4 | 31 | 92404  | 5  | 1.22 | 399.7 | 40 | 194907 |
| 6  | 0.03 | 394.6 | 18 | 4671   | 6  | 1.89 | 338.9 | 41 | 217073 |
| 7  | 1.00 | 344.9 | 39 | 118956 | 7  | 2.67 | 305.6 | 51 | 249355 |
| 8  | 1.09 | 326.1 | 42 | 115912 | 8  | 2.57 | 230.7 | 20 | 136782 |
| 9  | 1.77 | 303.9 | 26 | 163469 | 9  | 1.11 | 204.1 | 34 | 46239  |
| 10 | 0.70 | 257.6 | 37 | 46450  | 10 | 3.06 | 107.1 | 34 | 35099  |
| 11 | 0.50 | 225.9 | 56 | 25515  | 11 | 6.96 | 83.7  | 29 | 48760  |
| 12 | 0.10 | 206.7 | 61 | 4272   |    |      |       |    |        |
| 13 | 0.82 | 176.3 | 42 | 25487  |    |      |       |    |        |
| 14 | 0.28 | 130.9 | 18 | 4798   |    |      |       |    |        |
| 15 | 0.49 | 98.9  | 17 | 4793   |    |      |       |    |        |
| $\varepsilon_\infty = 5.50$ ||||| $\varepsilon_\infty = 3.33$ |||||

## Results of X-ray experiments

**Table SI7**. Lattice parameters and results of structural refinements.

| Parameters | $Pb_{0.93}Fe_{11.35}Zr_{0.65}O_{19}$ | $Ba_{0.2}Pb_{0.8}Fe_{12-x}Al_xO_{19}$ | | | |
|---|---|---|---|---|---|
| | | $x(Al) = 0.0$ | $x(Al) = 1.2$ | $x(Al) = 3.0$ | $x(Al) = 3.3$ |
| $T$ (K) | 100(1) | 296(2) ||||
| $a$ (Å) | 5.9036(1) | 5.89130(7) | 5.87110(6) | 5.8256(1) | 5.8228(4) |
| $c$ (Å) | 23.3611(5) | 23.1326(4) | 23.0277(4) | 22.8809(7) | 22.845(3) |
| $V$ (Å$^3$) | 705.11(3) | 695.31(2) | 687.42(2) | 672.49(4) | 670.79(12) |
| $R, F>2\sigma(F)$ | 0.0160 | 0.0277 | 0.0169 | 0.0188 | 0.0216 |
| Number of reflections: $N_{total}, N_{uniq}, R_{eqv}$ | 32008, 905, 0.044 | 44797, 1185, 0.036 | 44429, 1171, 0.028 | 37122, 1143, 0.028 | 43309, 1140, 0.043 |

**Table SI8**. Atomic coordinate parameters for tested compounds.

| Atom | | $Pb_{0.93}Fe_{11.35}Zr_{0.65}O_{19}$ | $Ba_{0.2}Pb_{0.8}Fe_{12-x}Al_xO_{19}$ | | | |
|---|---|---|---|---|---|---|
| | | | $x(Al) = 0.0$ | $x(Al) = 1.2$ | $x(Al) = 3.0$ | $x(Al) = 3.3$ |
| Pb1(Ba1) | $x$ | 0.63681(4) | 0.6506(4) | 0.64476(4) | 0.64822(6) | 0.64770(7) |
| | $y$ | 0.27362(8) | 0.3013(7) | 0.28953(8) | 0.29644(12) | 0.29541(13) |
| | $z$ | 0.24141(2) | 1/4 | 1/4 | 1/4 | 1/4 |
| Fe1(Al1) | $x$ | 0 | 0 | 0 | 0 | 0 |
| | $y$ | 0 | 0 | 0 | 0 | 0 |
| | $z$ | 0 | 0 | 0 | 0 | 0 |
| Fe2(Al2) | $x$ | 0 | 0 | 0 | 0 | 0 |
| | $y$ | 0 | 0 | 0 | 0 | 0 |
| | $z$ | 0.23797(3) | 0.24438(16) | 0.25611(8) | 0.25750(4) | 0.25765(5) |
| Fe3(Al3) | $x$ | 1/3 | 1/3 | 1/3 | 1/3 | 1/3 |
| | $y$ | 2/3 | 2/3 | 2/3 | 2/3 | 2/3 |
| | $z$ | 0.02660(2) | 0.02727(2) | 0.02749(2) | 0.02785(2) | 0.02793(2) |
| Fe4(Zr4/Al4) | $x$ | 1/3 | 1/3 | 1/3 | 1/3 | 1/3 |
| | $y$ | 2/3 | 2/3 | 2/3 | 2/3 | 2/3 |
| | $z$ | 0.18893(2) | 0.19028(2) | 0.19009(2) | 0.18960(2) | 0.18952(2) |
| Fe5(Al5) | $x$ | 0.16829(2) | 0.16876(2) | 0.16849(2) | 0.16831(2) | 0.16828(2) |
| | $y$ | 0.33658(4) | 0.33752(5) | 0.33697(3) | 0.33662(3) | 0.33657(5) |
| | $z$ | -0.10666(2) | -0.10860(2) | -0.10834(2) | -0.10804(2) | -0.10802(2) |
| O1 | $x$ | 0 | 0 | 0 | 0 | 0 |
| | $y$ | 0 | 0 | 0 | 0 | 0 |
| | $z$ | 0.14933(8) | 0.15094(10) | 0.15062(6) | 0.14931(6) | 0.14918(8) |
| O2 | $x$ | 1/3 | 1/3 | 1/3 | 1/3 | 1/3 |
| | $y$ | 2/3 | 2/3 | 2/3 | 2/3 | 2/3 |
| | $z$ | 0.55520(8) | -0.05491(10) | -0.05476(6) | -0.05468(6) | -0.05469(8) |
| O3 | $x$ | 0.18083(17) | 0.1816(2) | 0.18186(12) | 0.18205(12) | 0.18206(15) |
| | $y$ | 0.3617(3) | 0.3632(4) | 0.3637(2) | 0.3641(2) | 0.3641(3) |
| | $z$ | 1/4 | 1/4 | 1/4 | 1/4 | 1/4 |
| O4 | $x$ | 0.15596(11) | 0.15638(13) | 0.15428(8) | 0.15224(7) | 0.15191(9) |
| | $y$ | 0.3119(2) | 0.3128(3) | 0.30855(15) | 0.30447(15) | 0.30383(19) |
| | $z$ | 0.05211(5) | 0.05217(5) | 0.05147(3) | 0.05110(3) | 0.05103(4) |
| O5 | $x$ | 0.50399(11) | 0.50264(14) | 0.50273(9) | 0.50342(8) | 0.50349(11) |
| | $y$ | 0.0080(2) | 1.0053(3) | 1.00546(17) | 1.00685(16) | 1.0070(2) |
| | $z$ | 0.14887(4) | 0.14972(6) | 0.14906(3) | 0.14787(3) | 0.14768(4) |

**Table SI9.** Fe-O bond distances (Å); n – number of equivalent bonds.

| Cation | $Pb_{0.93}Fe_{11.35}Zr_{0.65}O_{19}$ | $Ba_{0.2}Pb_{0.8}Fe_{12-x}Al_xO_{19}$ | | | |
|---|---|---|---|---|---|
| | | $x(Al) = 0.0$ | $x(Al) = 1.2$ | $x(Al) = 3.0$ | $x(Al) = 3.3$ |
| | Fe-O bond distances (Å) | | | | |
| Fe1 | 2.0063(11), ×6 | 2.0006(13), ×6 | 1.9663(8), ×6 | 1.9305(7), ×6 | 1.9252(10), ×6 |
| Fe2 | 1.8703(17), ×3<br>2.071(2) | 1.858(2), ×3<br>2.161(4) | 1.8547(13), ×3<br>2.148(2) | 1.8449(12), ×3<br>2.1321(17) | 1.8444(15), ×3<br>2.128(2) |
| Fe3 | 1.9091(11), ×3<br>1.9110(19) | 1.8952(13), ×3<br>1.901(2) | 1.8938(14)<br>1.9028(8), ×3 | 1.8884(13)<br>1.9032(7), ×3 | 1.8874(17)<br>1.9043(10), ×3 |
| Fe4 | 1.9802(11), ×3<br>2.1135(13), ×3 | 1.9659(14), ×3<br>2.0749(17), ×3 | 1.9647(8), ×3<br>2.0678(9), ×3 | 1.9640(8), ×3<br>2.0592(9), ×3 | 1.9643(10), ×3<br>2.0582(12), ×3 |
| Fe5 | 1.9442(7), ×2<br>1.9886(10)<br>2.0720(11)<br>2.0939(8), ×2 | 1.9276(9), ×2<br>1.9811(12)<br>2.0887(14)<br>2.1137(9), ×2 | 1.9167(5), ×2<br>1.9706(7)<br>2.0816(8)<br>2.1033(6), ×2 | 1.8903(5), ×2<br>1.9432(7)<br>2.0648(8)<br>2.0814(5), ×2 | 1.8870(7), ×2<br>1.9404(9)<br>2.0627(10)<br>2.0790(7), ×2 |